\documentclass[twocolumn,showpacs,aps]{revtex4}
\usepackage{bm}
\usepackage{graphicx}
\usepackage{amsmath}
\usepackage{color}
\newcommand{\nix}[1]{}

\begin{document}

\title{Magnetic quantum ratchet effect in Si-MOSFETs}

\author{S.~D. Ganichev,$^1$ S.~A. Tarasenko,$^2$ J. Karch,$^1$ J. Kamann,$^1$ and Z.~D. Kvon$^3$}

\affiliation{$^1$Terahertz Center, University of Regensburg,
93040 Regensburg, Germany}
\affiliation{$^2$Ioffe Physical-Technical Institute, Russian Academy of
Sciences, 194021 St.\,Petersburg, Russia}
\affiliation{$^3$Institute of Semiconductor Physics, Russian Academy
of Sciences, 630090 Novosibirsk, Russia}

\begin{abstract}
We report on the observation of magnetic quantum ratchet effect in 
metal-oxide-semiconductor field-effect-transistors on silicon surface (Si-MOSFETs). 
We show that the excitation of an unbiased transistor by ac electric field of 
terahertz radiation at normal incidence leads to a direct electric current 
between the source and drain contacts if the transistor is subjected to an 
in-plane magnetic field. The current rises linearly with the magnetic field 
strength and quadratically with the ac electric field amplitude. It depends 
on the polarization state of the ac field and can be induced by both linearly 
and circularly polarized radiation. We present the quasi-classical and quantum 
theories of the observed effect and show that the current originates from the 
Lorentz force acting upon carriers in asymmetric inversion channels of the transistors.
\end{abstract}


\date{\today}

\maketitle

\section{Introduction}

A direct flow of charge carriers in semiconductor structures 
can be induced by $ac$ electric force with zero average driving. 
Such phenomenon, referred to as the electronic ratchet effect~\cite{Hanggi2009:RMP}, 
has been attracting much attention stimulated by both fundamental 
and applied interest in high-frequency non-linear electron transport 
at nanoscale~\cite{Linke1999:S, Entin2006, Chepelianskii2007, Sassine2008, Smirnov2008:PRL, Olbrich2010, Brizhik2010, Miyamoto2010, Roeling2011:NM, Kannan2012, Popov2013, Tanaka2013}. 
By its origin, the ratchet transport can occur only in structures with 
space inversion asymmetry and, therefore, 
it provides a powerful tool to study the symmetry properties of nanostructures, 
anisotropy of the band structure, electron-phonon and electron-impurity 
interactions and etc. Intrinsic and extrinsic ratchet mechanisms can awake 
the additional carrier degrees of freedom and drive 
spin~\cite{Costachel2010:N,Ganichev09} and 
valley~\cite{Karch11,Jiang13} currents. 
The application of an external static magnetic field breaks the time inversion 
symmetry giving rise to new mechanisms of current formation. In particular, it 
enables the magnetic quantum ratchet effect recently demonstrated for graphene 
layers excited by electromagnetic wave of terahertz (THz) range~\cite{Drexler13}.
The  effect emerges due to the joint orbital action of the $ac$ electric and static magnetic fields 
on two-dimensional (2D) electron gas in systems with structure 
inversion asymmetry (SIA). The latter leads to an asymmetric scattering of carriers 
in the momentum space resulting in a direct electric current~\cite{Falko89,Tarasenko11}.
While the SIA in graphene results from indistinct factors, such as adatoms on its surface,
it be obtained in a controllable way by application of a gate voltage to the electron channel in field-effect transistors.
Consequently, studying the magnetic quantum ratchet in such systems provides an access to the better understanding of this phenomenon.

Here, we report the observation and study of the magnetic quantum ratchet transport of electrons in Si-based field-effect-transistors 
(MOSFETs). 
We show that the excitation of the electron gas in the inversion channel of Si-MOSFETs subjected to an 
in-plane magnetic field by $ac$ electric field, here of THz radiation, leads 
to a direct electric current between the unbiased source and drain contacts. The 
current is proportional to the square of the ac electric field amplitude, scales 
linearly with the magnetic field strength, and reverses its direction by switching 
the magnetic field polarity. It can be generated by both linearly polarized and 
circularly polarized radiation. For linear polarization, the current depends on 
the angle between the electric field polarization and the static magnetic field. 
For circular polarization, the current reveals a helicity-sensitive component reversing 
its sign by switching the rotation direction of the electric field. We present the 
microscopic model of the effect as well as the quasi-classical and quantum theories 
explaining all major features observed in experiment. It is shown that the action 
of the Lorentz force on the electron motion induced by high-frequency electric field 
affects the electron scattering yielding an asymmetric distribution of nonequilibrium 
electrons.

\section{Samples and Technique}

We study $n$-type MOSFETs fabricated on (001)-oriented 
silicon surfaces by means of standard metal-oxide-semiconductor 
technology including preparation of SiO$_2$ with a thickness of 110~nm
by high temperature oxidation of silicon, preparation of heavily doped $n^{++}$ 
contacts by ion-implantation, and the fabrication of heavily doped polycrystalline 
semitransparent gates. Transistor with a channel 
length of $3$\,mm and  a width of $2.8$\,mm were 
prepared along $y \parallel [110]$. A doping level $N_a$ of the depletion 
layer was of about $3 \times 10^{15}$\,cm$^{-3}$. In these transistors, 
the variation of the gate voltage V$_g$ from $1$ to $20$\,V changes 
the carrier density $N_s$ from about $1.9 \times 10^{11}$ to $3.8 \times 10^{12}$\,cm$^{-2}$ 
and the energy spacing $\varepsilon_{21}$ between the size-quantized subbands $e1$ and 
$e2$ from $10$ to $35$\,meV. The peak mobilities $\mu$ at room and liquid 
helium temperature were 
$700$ and $10^4$\,cm$^2$/Vs, respectively.

To generate ratchet currents in unbiased samples we 
used alternating electric fields $\bm{E}(t)$ of a pulsed
terahertz NH$_3$  laser, 
optically pumped by a transversely excited atmosphere pressure (TEA) 
CO$_2$ laser~\cite{tunnelreview,PRL10}.
 The laser operated at frequencies $f=3.32$~THz (wavelength $\lambda = 90.5$~$\mu$m),
2.03~THz ($\lambda = 148$~$\mu$m) or 1.07~THz ($\lambda = 280$~$\mu$m).
It provides single pulses with a duration of about 100\,ns, peak power 
of $P \approx 30$\,kW, and a repetition rate of 1\,Hz. The radiation power was 
controlled by the THz photon drag detector~\cite{Ganichev84p20}.
The radiation at normal incidence was focused in a spot of 
about 1 to 3~mm diameter.  The spatial beam 
distribution had an almost Gaussian profile which was 
measured by a pyroelectric camera~\cite{Ziemann2000p3843}.
The experimental geometry is illustrated in the inset of Fig.~\ref{figure1}. 
All experiments are performed at room temperature and normal incidence of radiation.
In this geometry, the THz radiation causes \textit{intra}subband 
indirect optical transitions (Drude-like free carrier absorption).
An external in-plane magnetic field $\bm B$ of $\pm 1$~T was applied,
either parallel or perpendicular to the MOSFET channel, see insets of Fig.~\ref{figure1}.
The photocurrents are measured between the 
source and drain contacts of the unbiased transistors via the
voltage drop across a 50~$\Omega$ load resistor.

To vary the radiation polarization, $\lambda$/2 and $\lambda$/4 
crystal quartz plates were employed.
By applying the $\lambda/2$ plates, we varied the azimuth angle $\alpha$ 
between the polarization plane of the radiation incident upon the sample
and the $y$ axis. By applying $\lambda$/4 plates, we obtained 
elliptically (and circularly) polarized radiation. In this case, the polarization state 
is determined by the angle $\varphi$ 
between the plate optical axis and the incoming laser polarization with electric field vector along 
the $y$ axis. In particular, the radiation helicity is given by 
$P_{\rm circ} = \sin{2 \varphi}$~\cite{Ganichev02}.
The incident polarization states are sketched for characteristic angles $\varphi$ on top of
Fig.~\ref{figure2}.

\section{Experimental Results}
\label{sresults}

\begin{figure}[t]
\includegraphics[width=0.8\linewidth]{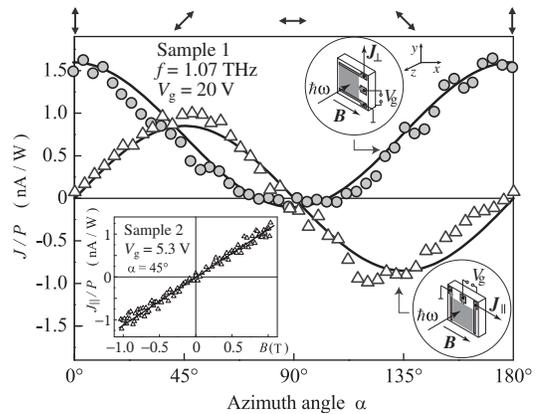}
\caption{Photocurrent as function of the azimuth angle 
$\alpha$ measured in 
sample~1 in the direction perpendicular and parallel to the magnetic field.
Here, a gate voltage of $V_{\rm g} = 20$\,V and radiation frequency $f=1.07$~THz
were used.
Lines are fits to Eqs.~(\ref{fitalpha}). 
The lower right inset shows the behavior of the parallel photocurrent $J_{\|}$ 
in sample 2 upon variation of the magnetic field strength. The dependence is obtained for
the azimuth angles $\alpha=45^\circ$ and $V_{\rm g} = 5.3$\,V.
The top and bottom left insets illustrate the experimental geometry.
The arrows on top illustrate the polarization states for different angles $\alpha$.
}
\label{figure1}
\end{figure}

Irradiating the transistors with polarized THz radiation we observed 
a photocurrent response, which scales linearly with the magnetic 
field strength and changes its sign by reversing the magnetic field direction, 
see inset in Fig.~\ref{figure1}. For zero magnetic field the photocurrent vanishes.
The measured electric current pulses are of about 100~ns duration and reflect 
the corresponding laser pulses.
The photocurrent is detected in the 
direction perpendicular to the magnetic field (transverse photocurrent, see 
top inset in Fig.~\ref{figure1}), as well as along $\bm{B}$ 
(longitudinal photocurrent, see bottom inset of Fig.~\ref{figure1}).  
These photocurrents exhibit characteristic polarization dependences
which are different for linear and elliptical polarized radiation. 

Figure~\ref{figure1} shows the dependence of the transverse and longitudinal photocurrents
on the \textit{ac} electric field azimuth angle $\alpha$ obtained for sample~1. 
The experimental data can be well fitted by 
\begin{equation} 
\label{fitalpha}
J_\bot (\alpha) = \chi_1 \cos{2 \alpha} + \chi_2 \:,\ \           
J_{\|} (\alpha) =  \chi_1 \sin{2 \alpha},
\end{equation}
where  $\chi_1$ and $\chi_2$ are fit parameters, see solid and dashed 
fit curves in Fig.~\ref{figure1}. Figure~\ref{figure1} and Eqs.~\eqref{fitalpha} 
reveal that the polarization dependent contributions to the transverse and longitudinal
photocurrents vary according to the Stokes parameters $S_1=\cos{2 \alpha}$ and $S_2=\sin{2 \alpha}$ 
multiplied by the same fit coefficient $\chi_1$. The polarization independent 
contribution is detected for the transverse photocurrent 
only and is described by the parameter $\chi_2$.

\begin{figure}[t]
\includegraphics[width=0.8\linewidth]{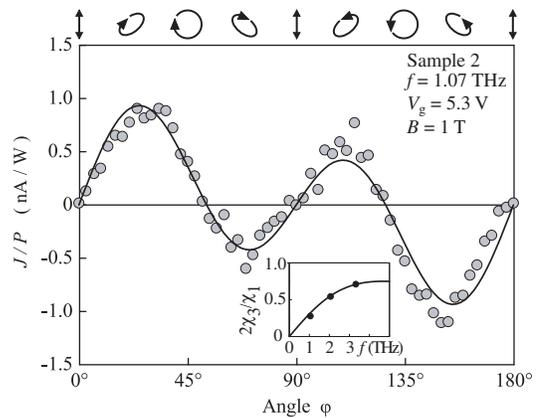}
\caption{Photocurrent as a function of the radiation helicity
measured in the direction parallel to the magnetic field.
The data are obtained for $V_{{\rm g}} = 5.3$\,V and
radiation with the frequency $f=1.07$~THz.
The line is a fit to the phenomenological Eq.~(\ref{fitphi}).
In the inset, the behavior of the ratio of magnetic 
field induced circular and linear photocurrents
upon variation of the wavelength is demonstrated.
On top the polarization ellipses corresponding to various angles
$\varphi$ are illustrated.
} \label{figure2}
\end{figure}

The photocurrent in response to elliptically polarized radiation is 
shown in Fig.~\ref{figure2}. 
In this case, the transverse and longitudinal
photocurrent components can be well fitted by
\begin{equation} \label{fitphi}
J_\bot  = \chi_1 \cos^2 2 \varphi + \chi_2 \:,\;\;
J_{\|}  = \frac{\chi_1}{2} \sin 4 \varphi + \chi_3 P_{\rm circ}  \:. 
\end{equation}
Obviously all current contributions measured for linearly polarized radiation
are detected in the experimental geometry applying a $\lambda/4$-plate.
These are the terms proportional to the fit parameters $\chi_1$ and $\chi_2$. 
The only change is the functional behavior of the Stokes parameters which now
are given by $S_1=\cos^2 2 \varphi$ and $S_2=\sin 4 \varphi/2$. 
In the longitudinal geometry, 
however, a new photocurrent contribution is observed. 
It is proportional to the 
radiation helicity $P_{{\rm circ}}$ corresponding to the third Stokes parameter $S_3$.
This contribution exhibits the sign inversion upon switching the radiation helicity $P_{{\rm circ}}$ from +1
to -1 at $\varphi = 45^\circ$ $(\sigma_+)$ and
$\varphi = 135^\circ$ $(\sigma_+)$, respectively. 
We note that, for pure circularly polarized radiation, the Stokes parameters $S_1$ and $S_2$ vanish.
Similar to the photocurrent induced by linearly polarized radiation,
the circular photocurrent linearly scales with the magnetic field strength (not shown).

Applying the radiation of various frequencies, we observed that the ratio $\chi_3 / \chi_1$,
which determines the relative magnitude of the circular and linear photocurrents,
increases with raising the radiation frequency $f$ at small $f$. The corresponding spectral dependence of 
$\chi_3 / \chi_1$ obtained for sample 2 is shown in the inset in Fig.~\ref{figure2}.

All the observations, including the linear dependence of the electric response on
 the static magnetic field $\bm B$, the square dependence on the radiation field 
amplitude as well as the polarization dependence, exhibit the recognized behavior 
of the magnetic quantum ratchet effect. Below, we present the misroscopic model 
and detail theory of the effect and show that they describe well the experimental data.

\begin{figure}[t]
\includegraphics[width=0.95\linewidth]{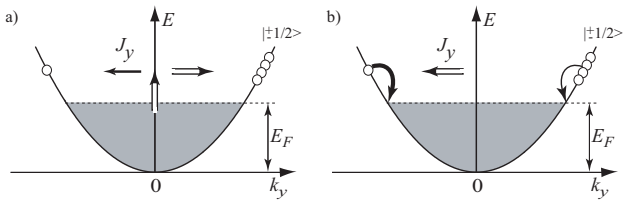}
\caption{Orbital mechanisms underlying the generation of the MPGE current.
(a) Excitation and (b) relaxation mechanisms.
} \label{Model}
\end{figure}

\section{Microscopic theory}

Microscopic mechanisms responsible for the observed photocurrent in Si-based 
structures involve asymmetry of the photoexcitation (excitation mechanism) or 
relaxation (relaxation mechanism). Both mechanisms are of a pure orbital origin 
and based on the asymmetry of electron scattering by static defects or phonons in 
the momentum space~\cite{Falko89,Kibis99,Tarasenko08}. The scattering asymmetry is 
caused by the Lorentz force acting upon carriers in inversion channels. It is 
describe by the correction to the scattering rate $W_{\bm{k}'\bm{k}}$ that is 
linear in the wave vector and in-plane magnetic field. Such a correction is 
allowed in gyrotropic structures only and, in Si-MOSFETs, is caused by structure 
inversion asymmetry of inversion channels. It can be obtained microscopically by 
considering the magnetic field induced change of the electron wave function, 
which yields~\cite{Tarasenko08}
\begin{equation}\label{W_asym}
W_{\bm{k}'\bm{k}} = W_{\bm{k}'\bm{k}}^{(0)} + w [B_x(k_y + k'_y) - B_y(k_x + k'_x)] \:,
\end{equation}
where $\bm{k}$ and $\bm{k}'$ are the initial and scattered wave vectors, 
$W_{\bm{k}'\bm{k}}^{(0)}$ is the scattering rate at $\bm{B}=0$, and the 
coefficient $w$ describes the scattering asymmetry degree. Due to linear in 
the wave vector terms in $W_{\bm{k}'\bm{k}}$, the scattering processes to the 
states $\bm{k}'$ and $-\bm{k}'$ occur at different probabilities, which results 
in asymmetric distribution of electrons in $\bm{k}$-space if the electron gas is 
driven out of equilibrium.
This is the origin of $dc$ electric current in Si-MOSFETs excited by terahertz radiation. 

The excitation and relaxation mechanisms of the current generation are 
illustrated in Figs.~\ref{Model}~(a) and~(b), respectively. Figure~\ref{Model}~(a)
sketches the intrasubband absorption of radiation (Drude absorption)
which includes a momentum transfer from phonons or impurities to electrons to 
satisfy momentum conservation. The vertical arrow shows electron-photon interaction 
while the horizontal
arrows describe the elastic scattering events to the final state with either positive 
or negative electron wave vector $k^\prime_x$. In the in-plane magnetic field, the 
probabilities of scattering to positive and negative $k^\prime_x$ are not equal, which 
is shown by the horizontal arrows of different thickness. This leads to an asymmetric 
distribution of photoexcited carriers in $\bm{k}$-space, i.e., an electric current $\bm{j}$. 
For simplicity,
we have drawn transitions only from the initial state $k_x=0$, however the argument holds 
for arbitrary $k_x$ as well.   
The electric current $\bm{j}$ is odd in the magnetic field $\bm{B}$ because the asymmetric 
part of scattering rate is proportional to $\bm{B}$, see Eq.~(\ref{W_asym}), and the probabilities 
for scattering to the positive or negative $k^\prime_x$ are inverted for the fields $\bm{B}$ and $-\bm{B}$.
For linearly polarized radiation, the momenta of photoexcited carriers are preferably aligned 
along the electric field of radiation. Therefore, for a fixed magnetic field, the current 
direction and magnitude depend on the radiation polarization state. 
The electric current caused by asymmetry of photoexcitation decays within the typical 
momentum relaxation time of electrons after the irradiation is switched off.    

Similarly to photoexcitation, the energy relaxation of hot carriers due to inelastic 
scattering by phonons is also asymmetric in $\bm{k}$-space, which leads to an additional 
contribution to the electric current. This relaxation mechanism is illustrated in 
Fig.~\ref{Model}~(b) where the curved arrows of different thickness show the inequality 
of relaxation rates at positive and negative $k_x$. The relaxation photocurrent is also 
odd in the magnetic field $\bm{B}$.
However, it is independent of the radiation polarization and decays within the energy 
relaxation time after the optical excitation pulse. Thus, photocurrent measurements with 
high time resolution can be used to distinguish between the excitation and relaxation mechanisms.

Below, we present quasi-classical and quantum theories of magnetic quantum ratchet 
effect. The quasi-classical approach is valid provided the photon 
energy $\hbar\omega$ is much smaller than the mean kinetic energy of carriers
 $\tilde{\varepsilon}$ and developed in the framework of Boltzmann's equation. 
The quantum theory is required if $\hbar\omega$ is comparable to or exceeds 
$\tilde{\varepsilon}$ and involves the quantum mechanical consideration of 
intrasubband optical transitions. For simplicity, we consider below the carriers 
in the ground subband of size quantization $e1$. The population of excited 
subbands at high temperature or small gate voltage may modify the current amplitude.

\subsection{Quasi-classical approach}

The quasi-classical theory of the magnetic quantum ratchet effect is developed following Refs.~\cite{Falko89,Tarasenko11}.
In this approach, the electric field of the radiation 
$\bm{E}(t)=\bm{E}\exp(-i\omega t)+\bm{E}^*\exp(i\omega t)$ is considered 
as {\it ac} force acting upon charge carriers. The electron distribution 
function $f_{\bm{k}}$ in $\bm{k}$-space is found from the Boltzmann equation 
\begin{equation}\label{kineq}
\frac{\partial f_{\bm{k}}}{\partial t} + e \bm{E}(t) \cdot \frac{\partial f_{\bm{k}}}{\hbar\, \partial \bm{k}} = {\rm St} f_{\bm{k}} \:,
\end{equation}
where ${\rm St} f_{\bm{k}}$ is the collision integral. For elastic scattering, 
${\rm St} f_{\bm{k}}$ has the form
\begin{equation}\label{St}
{\rm St} f_{\bm{k}} = \sum_{\bm{k}'} (W_{\bm{k}\bm{k}'} f_{\bm{k}'} - W_{\bm{k}'\bm{k}} f_{\bm{k}'} ) \:,
\end{equation}
where $W_{\bm{k}'\bm{k}}$ is the scattering rate. Taking into account the 
admixture of excited-subband states to the ground-subband wave function in the in-plane 
magnetic field $\bm{B}$, one obtains 
the matrix element of electron scattering~\cite{Tarasenko08}
\begin{equation}
V_{\bm{k}'\bm{k}} = V_{11} - \frac{e\hbar [B_x(k_y + k'_y) - B_y(k_x + k'_x)]}{m^* c} \sum_{\nu \neq 1} \frac{z_{\nu 1} V_{1\nu}}{\varepsilon_{\nu 1}} 
\end{equation}
and the scattering rate Eq.~(\ref{W_asym}) with the parameters 
\begin{equation}\label{W_0}
W_{\bm{k}'\bm{k}}^{(0)} = \frac{2\pi}{\hbar} |V_{11}|^2 \, \delta(\varepsilon_{\bm{k}}-\varepsilon_{\bm{k}'}) \:,
\end{equation}
\begin{equation}\label{w_correction}
w = - \frac{4 \pi e}{m^* c} \sum_{\nu \neq 1} \frac{z_{\nu 1}}{\varepsilon_{\nu 1}} \, {\rm Re} (V_{11}^* V_{1\nu}) \, \delta(\varepsilon_{\bm{k}}-\varepsilon_{\bm{k}'}) \:.
\end{equation}
Here, $V_{11}$ and $V_{1 \nu}$ ($\nu \neq 1$) are the matrix elements of intrasubband 
and intersubband scattering at $\bm{B}=0$, $\nu$ is the subband index, 
$\varepsilon_{\bm{k}} = \hbar^2 \bm{k}^2 /(2m^*)$, $m^*$ is the effective mass 
in the channel plane, $e$ is the electron charge, $z_{\nu 1}$ are the coordinate 
matrix elements, and $\varepsilon_{\nu 1}$ are the energy distances between the 
subbands. Note that, for Si-MOSFETs on the (001) surface, the in-plane mass $m^*$ 
is given by $m_{\perp}$ while the energies $\varepsilon_{\nu 1}$ are determined 
by $m_{\parallel}$, where $m_{\perp}$ and $m_{\parallel}$ are the transversal and 
longitudinal effective electron mass in bulk Si.

The electric current density is given by
\begin{equation}\label{current_gen}
\bm{j} = 4 e \sum_{\bm{k}} \bm{v}_{\bm{k}} f_{\bm{k}} \:,
\end{equation}
where $\bm{v}_{\bm{k}} = \hbar \bm{k}/m^*$ is the velocity and the factor~4 
accounts for the spin and valley degeneracy. By solving the Boltzmann 
equation~(\ref{kineq}) to second order in the electric field amplitude $\bm{E}$ 
and first order in the static magnetic field $\bm{B}$ one obtains $dc$ electric 
current. Calculation shows that the $y$ component of the current is given by 
\begin{equation}\label{j_MPGE}
j_y = (M_1 S_1 - M_2) |\bm{E}|^2 B_x  + (M_1 S_2 + M_3 S_3) |\bm{E}|^2 B_y  \:, 
\end{equation} 
where $S_1 = (|E_x|^2 - |E_y|^2)/|\bm{E}|^2$, $S_2 = (E_x E_y^* + E_y E_x^*)/|\bm{E}|^2$, and 
$S_3 = {\rm i} (E_x E_y^* - E_y E_x^*) /|\bm{E}|^2$ are the Stokes parameters determined 
by the radiation polarization, 
\begin{equation}\label{M1}
M_1 = \frac{\zeta \, e^4}{\pi m^* c \hbar^2} \int\limits_{0}^{\infty} \frac{ \tau_1 \, (\tau_1 \tau_2 \, \varepsilon^2)' f'_0 \, d\varepsilon}{1+(\omega \tau_1)^2}  \:, 
\end{equation}
\vspace{-0.5cm}
\begin{equation}\label{M2}
M_2 = \frac{\zeta \, e^4}{\pi m^* c \hbar^2} \int\limits_{0}^{\infty} \frac{(1-\omega^2 \tau_1 \tau_2) \tau_1 \tau_2 \, \varepsilon^2  \tau'_1 f'_0 \, d\varepsilon}{[1+(\omega \tau_1)^2][1+(\omega \tau_2)^2]}  \:, 
\end{equation}
\vspace{-0.5cm}
\begin{equation}\label{M3}
M_3 = - \frac{\zeta \, e^4}{\pi m^* c \hbar^2} \int\limits_{0}^{\infty} \frac{ \omega \tau_1 \tau_2 (\tau_1+\tau_2) \, \varepsilon^2 \tau'_1 f'_0 \, d\varepsilon }{[1+(\omega \tau_1)^2][1+(\omega \tau_2)^2]} \:,
\end{equation}
$\zeta = (4m^*/\hbar^3)\sum_{\nu \neq 1} z_{\nu 1} \rm{Re}(V_{11}^* V_{1\nu})/\varepsilon_{\nu 1}$,
$\tau_1$ and $\tau_2$ are respectively the relaxation times of the first and 
second angular harmonics of the distribution function in the absence of magnetic field,
\[
\tau_{n}^{-1} = \sum_{\bm{k}'} W_{\bm{k}\bm{k}'}^{(0)} (1-\cos n\theta_{\bm{k}'\bm{k}}) \:, 
\]
$\theta_{\bm{k}'\bm{k}}$ is the angle between the wave vectors
 $\bm{k}'$ and $\bm{k}$, $\tau'_1 = d \tau_1 / d \varepsilon$, $f'_0 = d f_0(\varepsilon) /d \varepsilon$, 
and $f_0(\varepsilon)$ is the function of equilibrium carrier distribution. 

Equations~(\ref{j_MPGE})-(\ref{M3}) describe the excitation mechanism of current formation.
The first and second terms on the right-hand side of Eq.~(\ref{j_MPGE}) stand for the 
current components perpendicular and parallel to the applied magnetic field, respectively. 
The perpendicular component contains the contribution sensitive to linear polarization of 
the radiation, $\propto S_1$, and the polarization independent term. The parallel component 
depends on both linear, $\propto S_2$, and circular, $\propto S_3$, polarization states of 
the radiation. 

At small frequencies of the ac electric field, the linear photocurrent given by $M_1$ is 
independent of $\omega$ while the circular photocurrent given by $M_3$ is proportional to $\omega$, 
as observed in the experiment. Considering the Boltzmann distribution of carriers, 
$f_0(\varepsilon) \propto \exp(-\varepsilon/k_B T)$, and the power dependence of the 
relaxation times on energy, 
$\tau_1(\varepsilon) = a \varepsilon^{r}$, $\tau_2(\varepsilon)/\tau_1(\varepsilon) = (2-r)/2$, 
one obtains  
\begin{equation}\label{ratio}
\frac{M_3}{M_1} = \omega \tau_1(k_B T) \frac{r (r-4)}{4(r+1)} \, \frac{\Gamma(4r+2)}{\Gamma(3r+2)}  \:,
\end{equation}
where $\Gamma(x)$ is the Gamma function. At high frequency, $\omega \gg 1/\tau_1$, 
the linear and circular currents decrease as
$1/\omega^2$ and $1/\omega^3$, respectively, and the linear current dominates.  

We note that that the energy relaxation of hot carriers in the in-plane magnetic field 
leads to an additional contribution to the polarization-independent current given by 
$M_2$, see Fig.~\ref{Model}(b). This contribution depends on the details of electron-phonon 
scattering and was theoretically addressed in Refs.~\cite{Kibis99,Tarasenko08}.

\subsection{Quantum approach}

The relevant description of the magnetic ratchet effect in the quantum 
regime involves the consideration of indirect intrasubband optical transitions. 
Due to energy and momentum conservation, the intrasubband absorption of radiation 
is accompanied by electron scattering from static defects or phonons. Such 
second-order processes are theoretically described by virtual transitions with 
intermediate states. 
Taking into account the virtual transitions via states in the ground and excited 
electron subbands one can obtain the matrix element of the intrasubband transitions 
$\bm{k} \rightarrow \bm{k}'$. To first order in the in-plane magnetic field, the
 matrix element of the intrasubband transitions accompanied by elastic electron 
scattering has the form
\begin{equation}
M_{\bm{k}'\bm{k}} = \frac{e \bm{A} \cdot (\bm{k}'-\bm{k})}{c \, \omega \, m^*} V_{\bm{k}'\bm{k}} - 2 \frac{e^2 (A_x B_y - A_y B_x)}{m^* c^2} \sum_{\nu \neq 1} \frac{z_{\nu 1}}{\varepsilon_{\nu 1}} V_{1 \nu} \:,
\end{equation}
where $\bm{A}$ is the amplitude of the electromagnetic field vector potential, $\bm{A}=(-i c/\omega)\bm{E}$.

The radiation absorption in the presence of the in-plane magnetic field 
leads to an asymmetry in the electron distribution in $\bm{k}$-space and, 
hence, to an electric current. The anisotropic part of the electron distribution 
function can be found from the master equation
\begin{equation}
g_{\bm{k}} = {\rm St} f_{\bm{k}} \:,
\end{equation}
where $g_{\bm{k}}$ is the generation rate due to intrasubband optical transitions
\begin{equation}
g_{\bm{k}} = \frac{2 \pi}{\hbar} \sum_{\bm{k}',\pm} |M_{\bm{k}'\bm{k}}|^2 [f_0(\varepsilon_{\bm{k}'}) - f_0(\varepsilon_{\bm{k}})] \, \delta(\varepsilon_{\bm{k}'} - \varepsilon_{\bm{k}} \pm \hbar\omega) .
\end{equation}

Taking into account linear-in-$\bm{B}$ terms both in the generation rate $g_{\bm{k}}$ 
and the collision integral ${\rm St} f_{\bm{k}}$ one can calculate the asymmetric part 
of the distribution function and the electric current. The electric current
originating from the asymmetric part of $g_{\bm{k}}$ contains both polarization-dependent 
and polarization-independent contributions. It is caused by asymmetry of optical 
transitions in $\bm{k}$-space and decays with the momentum relaxation time. The current 
stemming from the asymmetric part of ${\rm St} f_{\bm{k}}$ is polarization-dependent and 
vanishes for unpolarized radiation. It can be interpreted in terms of the optical alignment 
of electron momenta in $\bm{k}$-space by linearly polarized radiation followed by scattering 
asymmetry. Generally, the current contributions originating from the asymmetry of the 
generation rate and the collision integral are comparable to each other. For a particular 
case of short-range scattering, where the relaxation times of all non-zero angular harmonics 
of the distribution function coincide and are independent of energy, the electric current 
is given by Eq.~(\ref{j_MPGE}) with 
\begin{equation}\label{M1_quantum}
M_1 = - \frac{\zeta \, e^4 \, \tau_1}{\pi m^* c \hbar^3 \omega^3} \int\limits_{0}^{\infty} (2\varepsilon + \hbar\omega) [f_0(\varepsilon)- f_0(\varepsilon+\hbar\omega)] d\varepsilon \:. 
\end{equation}
Equation~(\ref{M1_quantum}) is valid for $\omega\tau_1 \gg 1 $ and can be considered 
as an extension of Eq.~(\ref{M1}) to high-frequency range. Naturally, both the 
quasi-classical and quantum approaches unite and yield the same result, compare 
Eqs.~(\ref{M1}) and~(\ref{M1_quantum}), for the intermediate frequency 
range $1/\tau_1 \ll \omega \ll \tilde{\varepsilon}/\hbar$.

\subsection{Estimation for triangular channel}

Before discussing experimental results in a view of the developed theory we estimate the ratchet current magnitude for Si-MOSFET structures. As follows from 
Eqs.~(\ref{j_MPGE}) and~(\ref{M1}), the polarization-dependent contribution to the current 
induced by linearly polarized radiation for $\omega\tau_1 \leq 1$ and short-range scattering is given by   
\begin{equation}\label{j_estimate}
j = \frac{4 e^4 \tau_1^2 N_s B E^2}{c \, m^{*2}}  \left| \sum_{\nu \neq 1} \frac{z_{\nu 1}}{\varepsilon_{\nu 1}} \, \xi_{\nu} \right| \:, 
\end{equation}
where $\xi_{\nu}=\rm{Re}(V_{11}^* V_{1\nu})/|V_{11}|^2$ and $N_s$ is the electron density. 
We assume that electrons in Si-MOSFET are confined in the triangular channel with the 
potential energy $U(z)=\infty$ for $z<0$ and $U(z)=|eF|z$ for $z>0$, where $F$ is the 
effective electric field. The electron wave functions and energies for the triangular 
channel have the form
\[
\varphi_{\nu}(z) = C_{\nu} \left(\frac{2 m_{\parallel} |eF|}{\hbar^2}\right)^{1/6} \hspace{-2mm} {\rm Ai} \left[ \left(\frac{2 m_{\parallel} |eF|}{\hbar^2}\right)^{1/3} \hspace{-2mm} z +  \lambda_{\nu} \right] \:, 
\]
\begin{equation}\label{Airy_function}
\varepsilon_{\nu} = - \left(\frac{\hbar^2|eF|^2}{2 m_{\parallel}}\right)^{1/3} \hspace{-2mm} \lambda_{\nu} \:,
\end{equation}
where $C_{\nu}$ are the normalization constants, ${\rm Ai}(z)$ is the Airy function, 
and $\lambda_{\nu}<0$ are the roots of the Airy function. The coordinate matrix elements 
and the energy distances between the subbands are respectively given by
\begin{equation}
z_{\nu1} = C_1 C_{\nu}  \left(\frac{\hbar^2}{2 m_{\parallel} |eF|}\right)^{1/3} \hspace{-2mm} \int_{0}^{\infty} \hspace{-2mm} {\rm Ai}(x+\lambda_{\nu}) x {\rm Ai}(x+\lambda_{1}) dx \:,
\end{equation}
\begin{equation}
\varepsilon_{\nu1}= \left(\frac{\hbar^2|eF|^2}{2 m_{\parallel}}\right)^{1/3} \hspace{-2mm} ( \lambda_1 - \lambda_{\nu}) \:.
\end{equation}
For the model of short-range scatterers uniformly distributed in the channel, which is also 
relevant for quasi-elastic electron scattering by acoustic phonons, the parameters $\xi_{\nu}$ 
have the form
\begin{equation}
\xi_{\nu} = \frac{\int \varphi_1^3(z) \varphi_{\nu}^3(z) dz}{\int \varphi_1^4(z) dz} = \frac{C_{\nu}}{C_1} \frac{\int {\rm Ai}^3(x+\lambda_1) {\rm Ai}(x+\lambda_{\nu})dx}{\int {\rm Ai}^4(x+\lambda_1) dx} \:.
\end{equation}
Therefore, we finally obtain
\begin{equation}\label{j_estimate2}
j = \frac{4 e^4 \tau_1^2 N_s B E^2}{c \, m^{*2} |eF|} |{\cal C}| \:, 
\end{equation}
where ${\cal C}$ is the dimensionless parameter, 

\begin{eqnarray}
C  &=& \sum_{\nu \neq 1} \frac{C_{\nu}^2}{\lambda_1-\lambda_{\nu}} \frac{\int_0^{\infty} {\rm Ai}^3(x+\lambda_1) {\rm Ai}(x+\lambda_{\nu})dx}{\int_0^{\infty} {\rm Ai}^4(x+\lambda_1) dx} \nonumber \\
&\times& \int_{0}^{\infty} {\rm Ai}(x+\lambda_{\nu}) x {\rm Ai}(x+\lambda_{1}) dx \approx -0.085 \:. 
\end{eqnarray}
Note, that the amplitude $E$ of the electric field acting upon the carriers in the sample 
is related to the laser radiation intensity $I$ by $E=t\sqrt{2\pi I/c}$, where ${t=2/(n_{\omega}+1)}$
 is the amplitude transmission coefficient for the normally incident radiation and $n_{\omega}$ 
is the refractive index.

\section{Discussion}

The microscopic theory of the magnetic quantum ratchet effect presented above 
describes all major features observed in the experiment. First, it shows that 
the direct electric current is proportional to the square of the ac electric 
field amplitude $\bm E$, i.e., proportional to the radiation intensity $I$, scales 
linearly with the in-plane magnetic field $\bm B$ and reverses its direction by changing 
the magnetic field polarity, see Eq.~\eqref{j_MPGE}. Such a behavior is observed in the 
experiment. Second, Eq.~\eqref{j_MPGE} demonstrates that the current components perpendicular 
to and along the static magnetic field has different polarization dependence: the   
perpendicular component contains the contribution determined by the Stokes parameter $S_1$ 
and polarization independent term while the
parallel component depends on both linear, $\propto S_2$, and circular, $\propto S_3$, polarization 
states of the radiation. Exactly this polarization behavior of the photocurrent is observed in the 
experiment, see Figs.~\ref{figure1} and~\ref{figure2} and the empirical fit Eqs.~(\ref{fitalpha}) 
and~(\ref{fitphi}). Third, according to Eqs.~\eqref{ratio} the ratio of the linear to circular 
contributions to the ratchet current is proportional to the radiation frequency $\omega$ at 
$\omega \tau_1 < 1$, which also corresponds to the experimental data, see inset in Fig.~\ref{figure2}. 
Finally, we estimate the ratchet current magnitude for Si-MOSFETs following Eq.~\eqref{j_estimate2}.
The estimation of the current density $j$ normalized by the laser radiation intensity 
$I$ yields ${j/I \sim 1\times10^{-9}\,}$A\,cm/W for the electron density 
$N_s = 3.8 \times 10^{12}$\,cm$^{-2}$ and effective field $F = 1.4 \times 10^5$\,V/cm 
(corresponding to the energy distance $\varepsilon_{21}=35$\,meV at the gate voltage $V_g = 20$\,V, 
see Fig.~\ref{figure1}), relaxation time $\tau_1=0.5\times 10^{-13}$\,s determined from the 
room-temperature mobility $400$\,cm$^2$/V\,s, magnetic field $B=1$\,T, the effective in-plane 
mass $m^*\approx 0.2 m_0$ and the longitudinal mass in the valley $m_{\parallel} \approx 0.92 m_0$, 
with $m_0$ being the free electron mass, and silicon refractive index of silicon $n_{\omega}\approx3.4$. 
For the $3\times3.5$\,mm$^2$ transistor and the similar laser beam cros-section, 
it gives $J/P \sim 3 \times 10^{-9}$A/W, which is close to the signal magnitude detected in 
the experiment, see Fig.~\ref{figure1}. Thus, the simple model of the inversion channel with 
no fit parameters provides the qualitatively correct magnitude of the ratchet current.

\section{Summary}

To summarize, we have demonstrated the room-temperature magnetic quantum ratchet effect 
in Si-based metal-oxide-semiconductor field-effect-transistors with electron inversion 
channels. The direct current between the source and drain contacts of the unbiased 
transistor is excited by ac electric field with zero average driving in the presence 
of the static in-plane magnetic field. The effect is caused by orbital effects of the 
electric and field on two-dimensional electron gas confined in an asymmetric channel. 
It could be employed for designing the fast polarization-resolved Si-based detectors 
of terahertz and microwave radiation.

\acknowledgments Support from DFG (SFB~689), Linkage Grant of IB of BMBF at DLR, RFBR, and RF
President Grant MD-3098.2014.2 is gratefully acknowledged.

\end{document}